\documentclass[nofootinbib,prd,twocolumn,preprintnumbers,amsmath,amssymb]{revtex4-1}

\usepackage[normalem]{ulem}
\usepackage{times}
\usepackage{natbib}
\usepackage{amssymb,amsbsy,amsmath,amsfonts}
\usepackage{graphicx}
\usepackage{epsf,epsfig,float,latexsym,amsthm,fancyhdr,rotating}
\usepackage{graphics,psfrag,longtable}
\usepackage[svgnames]{xcolor}
\usepackage{ulem}
\usepackage{slashed,wasysym}
\usepackage[colorlinks]{hyperref}

\newcommand{\ca}{{\cal A}}
\newcommand{\co}{{\cal O}}
\newcommand{\ch}{{\cal H}}
\newcommand{\crr}{{\cal R}}
\newcommand{\cu}{{\cal U}}

\newcommand{\nn}{\nonumber}
\newcommand{\Ima}{{\rm Im}\,}

\newcommand{\be}{\begin{equation}}
\newcommand{\ee}{\end{equation}}
\newcommand{\beq}{\begin{equation}}
\newcommand{\eeq}{\end{equation}}
\newcommand{\bea}{\begin{eqnarray}}
\newcommand{\eea}{\end{eqnarray}}
\newcommand{\ba}{\begin{eqnarray}}
\newcommand{\ea}{\end{eqnarray}}

\begin{document}
\title{Probabilistic interpretation of compositeness relation for resonances}

\author{Zhi-Hui Guo$^{1,2}$}
\author{J.~A.~Oller$^{3}$}

\affiliation{
$^1$
Department of Physics, Hebei Normal University, Shijiazhuang 050024, People's Republic of China 
\\
$^2$ State Key Laboratory of Theoretical Physics, Institute of
Theoretical Physics, CAS, Beijing 100190, People's Republic of China
\\
$^3$ Departamento de F\'isica, Universidad de Murcia,  E-30071 Murcia, Spain
}

\begin{abstract}
Bound, antibound and resonance states are associated to poles in the on-shell partial wave amplitudes. 
 We show here that from the residues of the pole  a rank 1 projection operator associated with any of these  states can be extracted, 
 in terms of which a sum rule related to the composition of the state can be derived. 
 Although typically it involves complex coefficients for the compositeness and elementariness, except for the bound state case,
   we demonstrate  that one can  formulate  a meaningful compositeness relation with only positive coefficients for resonances
 whose associated Laurent series in the variable $s$ converges in a region of the physical axis around $ {\rm Re}{s_P}$, with 
$s_P$ the pole position of the resonance.
 It is also shown that this result can be considered as an analytical extrapolation   in $s_P$  of the clear narrow resonance case. 
We exemplify this formalism to study the two-body components of several resonances of interest. 

\end{abstract}
\maketitle

\section{Introduction}

Resonance, as it arises in quantum-mechanical problems, is a common phenomenon in several branches of physics, e.g. 
in particle, nuclear, atomic and condensed matter physics. Here we make use of $S$ matrix theory, 
but the same principles hold in  other disciplines as well. An interesting question in spectroscopy is to 
understand the nature of resonances, so that one can see whether they can fit within the standard model of physics or they 
have another origin. Even in the former case, the situation requires  further clarification in many instances 
 in order to understand their nature in terms of the appropriate degrees of freedom. 

During the past decades there has been an increasing growth of evidences that hadron resonances do not always fit 
within the standard picture for mesons ($q\bar{q}$) and baryons ($qqq$) in quantum chromodynamics. Well known 
examples are  the early puzzles of the $\Lambda(1405)$ and related resonances, 
the lightest scalar mesons,  and the examples of new   
 bottomonium, charmonium  and charm-strange mesons with unexpected 
properties. In this respect, one also has the exciting new discovery of the 
$P_c(4450)$ \cite{ref.180815.1} that would require five valence quarks.  All these examples clearly show 
that the interest on hadron resonances and their nature is actually reinvigorated.

 A tool to understand the 
composition of a state is through the field renormalization constant $Z$,  first formulated for shallow
 bound states \cite{ref.040815.5}, and that represents the 
amount of extra components beyond the explicit degrees of freedom considered ($1-Z$ is then called the  
compositeness). 
 Its generalization to resonances  lying nearby a threshold is discussed in Refs.~\cite{ref.040815.6,ref.180815.2}.
 We focus here on another approach based on the analytical continuation 
of the compositeness relation to the pole position \cite{ref.040815.2,ref.040815.3,ref.040815.4}. This approach allows a probabilistic
 interpretation for two-body composition of bound states, but for the case of resonances its straightforward application 
drives to complex numbers which have prevented such interpretation. We show here how this difficulty can be overcome  
 by employing appropriate transformations driving to a new unitary $S$ matrix that shares the same resonant 
behavior but gives rise to compositeness relation involving only positive numbers.

\section{Sum rule}
\label{sec.260316.2}
 In the following we consider on-shell  two-body scattering of $n$ channels,
 and assume rotational and time reversal symmetry, so that 
the $S$ and $T$ matrices are symmetric  \cite{ref.300715.1}.
The $T$ matrix  can be written in general form  as \cite{ref.040815.1} 
\begin{align}
\label{040815.2}
T(s)=&\left[{\cal K}(s)^{-1}+G(s)
\right]^{-1}~,
\end{align}
where $s$ is the  usual Mandelstam variable and $G(s)$ is a diagonal matrix 
with matrix elements 
 $\delta_{ij}G(s)_i$. The $G(s)_i$ are the unitarity scalar-loop functions that encode the 
two-body unitarity requirement, ${\rm Im}T^{-1}(s)_{ij}=-\delta_{ij}\theta(s-s_i) \rho(s)_i$,   as well as 
its analytical properties. These functions can be expressed as  \cite{ref.130815.2}
\begin{align}
\label{040815.3}
G(s)_i=&a(s_0)_i-\frac{s-s_0}{\pi}\int_{s_i}^\infty ds'\frac{\rho(s')_i}{(s'-s)(s'-s_0)}~,
\end{align} 
and $s_i$ is the threshold for the $i_{\rm{th}}$ channel and we order the states such that $s_i<s_{i+1}$. 
 In the previous expressions $\rho(s)_i$ is the relativistic invariant phase space factor for channel $i$, 
$\rho(s)_i=q(s)_i/8\pi\sqrt{s}$, with $q(s)_i$ the center-of-mass three-momentum. We denote $ \rho(s)$ as the
  diagonal matrix, with matrix elements  $\delta_{ij}\rho(s)_i$. 
 In addition, the matrix  ${\cal K}(s)$ in Eq.~\eqref{040815.2} includes all other contributions 
not arising from the two-body unitarity, such as the contact interactions and the crossed-channel effects.  
 In our notation the relation between the $S$ and $T$ matrices is 
\begin{align}
\label{300715.2}
S(s)=&I + i  (2\rho(s))^{\frac{1}{2}}T(s)(2\rho(s))^{\frac{1}{2}}~,
\end{align}
with $I$ the identity matrix.  Equation~\eqref{300715.2} can also be applied (through analytical continuation) 
even if $s$ is below the threshold for some of the channels because its restriction to the subspace of open channels gives the correct $S$ matrix. 

Next we assume that the $T$ matrix has a pole with a given set of quantum numbers in the appropriate Riemann sheet (RS) at $s_P$. 
This pole is called a bound state if its lies in the physical RS, $s_P\in \mathbb{R}$ and $s_P<s_1$;  
an antibound state if $s_P\in \mathbb{R}$, $s_P<s_1$ but it lies in the unphysical RS adjacent to the physical one for $s\gtrsim s_1$; 
and a resonance state in any other case  (in the following we denote all of them  as pole states).
 The residues of the $T$ matrix at $s_P$ are given by 
\begin{align}
\label{300715.3}
\lim_{s\to s_P}(s-s_P)T(s)=-\gamma \gamma^T~,
\end{align}
where $\gamma$ is an $n$ row vector, $\gamma^T=(\gamma_1,\ldots,\gamma_n)$, with   $\gamma_i$ typically complex 
numbers. Now, by employing the 
algebraic result that $dB(s)^{-1}/ds=-B^{-1}(s)dB(s)/ds B^{-1}(s)$, with $B(s)$ an invertible matrix, we take the derivative with 
respect to $s$ of both sides of Eq.~\eqref{040815.2} at $s=s_P$. From the double-pole term at $s_P$ it follows that 
\begin{align}
\label{040815.3b}
\gamma\gamma^T\!=\!-\gamma\gamma^T \left[\frac{dG(s_P)}{ds}-{\cal K}^{-1}\frac{d{\cal K}(s_P)}{ds}{\cal K}^{-1}\right]\gamma\gamma^T.
\end{align}
 Let us stress that the derivative of $G(s)_i$ is independent of the subtraction constant $a(s_0)_i$
 and subtraction point $s_0$.  It corresponds to a convergent three-point one-loop function with unit vertices \cite{fif0}.  
 Next, we multiply Eq.~\eqref{040815.3b} to the left by $\gamma^\dagger$ and to the right by $\gamma^*$,
 and as  $\gamma^\dagger\gamma \neq 0$ it can be simplified.  
Furthermore, in the last term on the right-hand side (rhs) of Eq.~\eqref{040815.3b}
  we  employ  that $\gamma= -{\cal K}G\gamma$, as follows by rewriting Eq.~\eqref{040815.2} as 
$T={\cal K}-{\cal K}GT$ and taking the limit $s\to s_P$. Then,  one obtains the sum rule (SR)
\begin{align}
\label{040815.5}
1=& - \gamma^T\frac{dG(s_P)}{ds}\gamma + \gamma^T G(s_P)\frac{d{\cal K}(s_P)}{ds}G(s_P)\gamma\\
=&\!\!\sum_{i,j=1}^n \!\!\left(
-\delta_{ij}\gamma_i^2 \frac{dG(s_P)_i}{ds}
+\gamma_i G(s_P)_i \frac{d{\cal K}(s_P)_{ij}}{ds} G(s_P)_j \gamma_j \right).\nn
\end{align}
This SR was already derived in the literature \cite{ref.040815.2,ref.040815.3,ref.040815.4}. 
 The partial compositeness coefficients from Eq.~\eqref{040815.5} are given by
 $X_i=-\gamma_i^2 dG(s_P)_i/ds$, their sum is the total 
compositeness, $X=\sum_{i=1}^n X_i$,  and the elementariness (the analytically continued field renormalization constant
 at the pole), 
$Z=1-X$, corresponds to the sum over the last term on the rhs of Eq.~\eqref{040815.5}. 
 For bound states  all these coefficients are positive and their connection with the different components in the 
properly normalized bound state is understood 
\cite{ref.040815.5,ref.040815.6,ref.040815.4}.\footnote{ There is also another proposal 
in the literature for higher partial waves \cite{ref.040815.3} that consists of using  
a  redefined $T$ matrix in order to get the SR. However, this is physically equivalent, 
and leads to the same results, both on the physical axis and for the values of $X_i$, 
by properly defining  the accompanying matrices  ${\cal K}(s)$ and $G(s)$ \cite{ref.060815.1}. }  
However, for the rest of the pole states Eq.~\eqref{040815.5} does not 
allow a straightforward statistical quantum mechanical interpretation 
 since the compositeness and elementariness coefficients become usually complex.

\subsection{Projection operator of rank 1} 
\label{sec.090316.1}

 Around the pole position we can write for 
the partial-wave projected $S$ matrix the Laurent expansion
\begin{align}
\label{300715.1}
S(s)=&\frac{\crr}{s-s_P}+S_0(s)~,
\end{align}
where   ${\cal R}$ is the matrix of residues and $S_0(s)$ comprises the nonresonant terms. 
From Eqs.~\eqref{300715.2} and \eqref{300715.3} one has the following relation between the residues of the $S$ and $T$ matrices,
\begin{align}
 \label{rst}
\crr=-2i\rho(s_P)^\frac{1}{2}\gamma\gamma^T \rho(s_P)^\frac{1}{2}~. 
\end{align}
  Let us show next that the multiplicative structure of the residues  implies the
 existence of a projection operator of rank 1. For that we rewrite the matrix $\crr$
 in an explicitly symmetric form as  
\begin{align}
\label{300715.4}
{\crr}=&i\lambda \co \ca \co ^T~,
\end{align}
with $\lambda$ a real normalization constant. 
 Combining Eqs.~\eqref{rst} and \eqref{300715.4} leads to 
\begin{align}
\label{040815.6}
\ca=&-\frac{1}{\lambda}\omega \omega^T~, 
\end{align}
 where
\begin{align}
\label{250316.4}
\omega=&\co^{-1}\widetilde{\gamma}~, \\
\widetilde{\gamma}=&(2\rho(s_P))^{\frac{1}{2}} \gamma~.\nn
\end{align} 
  Next we require that $\ca$ is a projection operator, so that $\ca^\dagger=\ca$ and $\ca^2=\ca$. 
 The former condition implies that the vector $\omega$ must be real or purely  
imaginary 
 (these are $n$ constraints on the $2n^2$ free parameters present in $\co$), and we take it real 
in the following (this can always be done because $\ca$ depends quadratically on $\omega$). 
Next, the latter condition fixes  $\lambda$ to 
\begin{align}
\label{260316.1}
 \lambda=-\omega^T \omega<0~.
\end{align} 
 This result also implies that $\ca$ is of rank 1 because 
its trace is 
\begin{align}
\label{260316.2}
\rm{Tr} \ca=-\frac{1}{\lambda}Tr \omega \omega^T=1~.
\end{align} 
Notice that given a projection $\ca$ with the above properties and $L$ a real orthogonal 
matrix then $L\ca L^T$ shares the same properties as $\ca$. 
  
A deeper understanding of the SR in Eq.~\eqref{040815.5} can be 
reached  by rewriting $\gamma\gamma^T$ 
in Eq.~\eqref{040815.3b} in terms of $\ca$ according to  Eq.~\eqref{040815.6}. 
 After simplifying  common factors  one has 
\begin{align}
\label{040815.4}
\ca=&
\frac{\lambda}{2} \ca \co^T \rho(s_P)^{-\frac{1}{2}} \left[\frac{dG(s_P)}{ds}-G(s_P)\frac{d{\cal K}(s_P)}{ds}G(s_P)\right] \nn\\
\times &
  \rho(s_P)^{-\frac{1}{2}} \co \ca~.
\end{align}
 Next, we take the trace on both sides of this equation,
  employ on the rhs the cyclic property of the trace and then Eq.~\eqref{040815.5} results. 
 Although the connection of this SR with the proper normalization of bound states was already derived \cite{ref.040815.5,ref.040815.6},
 now we understand that it is also equal to 1 for resonances and antibound states because of the proper normalization 
  (to 1) of the pole state associated with the rank 1 projection operator $\ca$.

\subsection{Open and closed channels}  

Let us follow the convention to write a hat  when denoting submatrices in the reduced space with $m$ open channels ($s_m<s<s_{m+1}$) and, 
when indicated, the subscript $m$ corresponds to the number of them. 
  Since again $\hat{ S}_m(s)$ is a unitarity matrix it has its associated $T$  matrix, $\hat{T}_m(s)$,
 which is the restriction of the 
original one in the subspace of open channels, and its residues  are denoted by $\hat{\gamma}$. The matrix 
  $\hat{T}_m(s)$  can be written analogously to 
Eq.~\eqref{040815.2} in terms of an effective  $\hat{{\cal K}}_m(s)$, which can be calculated from 
the original ${\cal K}(s)$ by integrating out heavier channels \cite{ref.040815.4}.
 We can also derive the subsequent SR in the reduced subspace, which reads
\begin{align}
\label{020715.9}
1=&\hat{\gamma}^T \left[-\frac{d\hat{G}_m(s_P)}{ds} + \hat{G}_m(s_P) \frac{d\hat{{\cal K}}_m(s_P)}{ds}\hat{G}_m(s_P)\right]\hat{\gamma}.
\end{align}
This new SR can be related to the original one, expressed for $n$ channels, 
 by introducing  an $m\times n$ matrix ${\cal H}$ such that 
\begin{align}
\label{020715.10}
{\cal H}=&\left(
\begin{array}{c|c}
\hat{I}_{m}& 0 
\end{array}
\right)~,\\
\label{020715.12}
\hat{T}_m(s)=&-\frac{i}{2}\hat{\rho}(s)^{-\frac{1}{2}} \left[ {\cal H} S(s){{\cal H}}^T - \hat{I}_m \right]  \hat{\rho}(s)^{-\frac{1}{2}}~,
\end{align}
 where  $\hat{G}_m(s_P)=\ch G(s)$ and $\hat{\gamma}=\ch \gamma$, with  $\hat{I}_m$ the $m\times m$ identity matrix.
As done above, the SR results by taking the derivative of 
Eq.~\eqref{020715.12} at $s_P$,
\begin{align}
\label{050815.3}
&\hat{\gamma}\hat{\gamma}^T \left[\hat{G}_m(s_P)\frac{d\hat{{\cal K}}_m(s_P)}{ds}\hat{G}_m(s_P)-\frac{d\hat{G}_m(s_P)}{ds}\right]
\hat{\gamma}\hat{\gamma}^T\nn\\
=&\hat{\gamma}\gamma^T \left[G(s_P)\frac{d{\cal K}(s_P)}{ds}G(s_P)-\frac{dG(s_P)}{ds}\right] \gamma\hat{\gamma}^T,
\end{align}
and  multiplying next both sides to the left by $\hat{\gamma}^\dagger$ and to the right by $\hat{\gamma}^*$. 
 As the product $\hat{\gamma}^\dagger\hat{\gamma}\neq 0$ 
 the equality between the SRs of Eqs.~\eqref{040815.5} and \eqref{020715.9} follows. 
  Note that since the $X_i$, $i=1,\ldots,m$, are  the same in both schemes, from the equality between 
 these SRs  one obtains the relation 
\begin{align}
\label{020715.21}
\hat{Z}=&\hat{\gamma}^T \hat{G}_m \frac{d\hat{{\cal K}}_m(s_P)}{ds}\hat{G}_m\hat{\gamma}
=Z+\!\!\!\sum_{i=m+1}^n X_i.
\end{align}
 Thus, the elementariness $\hat{Z}$  in the reduced space  contains also 
the contribution from two-body heavier channels \cite{ref.040815.4}. This
 result  is derived here in a fully general manner without referring to any specific interaction.

\section{Transformed sum rules}
\label{sec.260316.3}

We point out that the SR in Eq.~\eqref{020715.9} is not unique. Given an $m\times m$ unitary matrix $\hat{\cu}_m$   we define the transformation 
\begin{align}
\label{050815.1}
\hat{S}_u(s)=\hat{\cu}_m \hat{S}_m(s) \hat{\cu}_m^T~.
\end{align}
This $S$ matrix is symmetric and also unitary for $s_m<s<s_{m+1}$, 
  so that one can  identify a transformed $T$ matrix, 
$\hat{T}_u(s)$ (given by the rhs of Eq.~\eqref{020715.12} replacing $\ch\to \hat{\cu}_m$ and 
$S\to \hat{S}_m$),  which could be expressed as in Eq.~\eqref{040815.2} in terms of  a new $\hat{{\cal K}}_u(s)$. 
The matrix of residues for $\hat{S}_u(s)$ is $\hat{\crr}_u=\hat{\cu}_m \hat{\crr} \hat{\cu}_m^T$ or, 
in other words, the matrix $\hat{\co}$, cf. Eq.~\eqref{300715.4}, 
transforms to $\hat{\co}_u=\hat{\cu}_m \hat{\co} \hat{L}^T$,
 while the projection operator $\hat{\ca}_m$
 does as $\hat{L}\hat{\ca}_m \hat{L}^T$ (with $\hat{L}$ the aforementioned real orthogonal matrix).
  In terms of  the residues of $\hat{T}_u(s)$, 
  $\hat{\gamma}_u=\hat{\rho}(s_P)^{-\frac{1}{2}}\hat{\cu}_m \hat{\rho}(s_P)^{\frac{1}{2}} \hat{\gamma}$, 
and $\hat{{\cal K}}_u(s)$ 
 a new SR attached to $\hat{\ca}_m$  can be derived
\begin{align}
\label{050815.2}
1=&-\hat{\gamma}_u^T \frac{d\hat{G}_m(s_P)}{ds}\hat{\gamma}_u 
+ \hat{\gamma}_u^T \hat{G}_m \frac{d\hat{{\cal K}}_u(s_P)}{ds}\hat{G}_m\hat{\gamma}_u~.
\end{align}

\subsection{New compositeness relation}
\label{sec.260316.4}

 In the subsequent all our results are based on the hypothesis that  the pole $s_P$ lies in an unphysical RS adjacent 
to the physical RS in an interval $s_m<s<s_{m+1}$, such that $s_m<{\rm Re} s_P<s_{m+1}$. We define the resonance mass  squared,  $M_R^2$, as $M_R^2={\rm Re}s_P$. The important point  is that, as a result,  
 the convergence region of the Laurent series  
in Eq.~\eqref{300715.1} includes a region of the real axis around $s=M_R^2$ 
(since this point is closer to $s_P$ than any of the nearest thresholds $s_m$ or $s_{m+1}$.)\footnote{For resonances lying far enough in  the complex plane  it is not excluded that the Laurent series 
does not actually converge in this energy region because of the closeness of some singularity due to crossed channel dynamics. This depends 
on the particular dynamics affecting every independent process and should  be studied case by case.}
 We denote this working assumption as 
${\bf I}$. In particular it excludes from our considerations the case of antibound states or resonances that lie in the RS mentioned but with  
 $M_R^2<s_m$ or $M_R^2>s_{m+1}$  (e.g. this is the case of the $a_0(980)$ in   Refs.~\cite{ref.120815.3,ref.130815.2}.  
These types of resonant signals are an admixture
 between a pole and an enhanced cusp effect by the pole itself \cite{ref.100815.1}).
  Condition {\bf I} is certainly satisfied by  a narrow resonance.  
  But we should stress that ${\bf I}$ goes far beyond the narrow resonance case  (the latter is discussed 
in detail below in Sec.~\ref{sec.260316.5}). 

 Based on the condition ${\bf I}$ we can further inquire about the matrix $\co$ in Eq.~\eqref{300715.4}.
 Since we are going to make use of  the 
unitarity of the $S$ matrix around $s=M_R^2$,  we consider its restriction $\hat{S}_m(s)$  in the subset of the $m$ 
open channels. Correspondingly, the analysis undertaken above from Eq.~\eqref{300715.4} until concluding the existence of the 
 rank 1 projection operator is restricted to this subset  as well. 
Performing a transformation of the type of Eq.~\eqref{050815.1}  we would have
\begin{align}
\label{140815.1}
\hat{S}_u(s)=&\hat{\cu}_m\hat{S}_m(s) \hat{\cu}_m^T
=\frac{i\lambda  \hat{\ca}_m}{s-s_P}+\hat{\cu}_m\hat{S}_0(s) \hat{\cu}_m^T~,
\end{align}
where 
  $i\lambda \hat{\ca}_m=\hat{\cu}_m \hat{\crr}_m \hat{\cu}_m^T$. 
 As both matrices $\hat{\crr}_m$ and $\hat{\ca}_m$ are constant 
  we consider only constant $\hat{\cu}_m$.  
 This is why we disregarded any energy dependence 
for the transformation matrix in Eq.~\eqref{050815.1}.

    Now, the point is that  because condition  {\bf I} holds then
  we can move from the real axis onto the pole position (due to the convergence of the Laurent series for $s$ around $M_R^2$) and 
identify the matrix $\co$ with $\,{\hat{\cu}_m}^{-1}$. 
 Since it is a unitary transformation the real vector $\hat{\omega}$,
 given by $\hat{\omega}=\hat{\cu}_m  (2\rho(s_P))^\frac{1}{2} 
\hat{\gamma}$,
 must have a modulus squared equal to
\begin{align}
\label{250316.5} 
\hat{\omega}^T\hat{\omega}=\sum_{i=1}^m|\widetilde{\gamma_i}|^2~.
\end{align} 
 This constraint cannot be determined from the analysis undergone above, which concluded with the existence of the 
projection operator $\ca$, because there we only attended to the factorizing structure of the residues, 
 while here we assume that {\bf I} holds.  Indeed, given a real vector $\hat{\omega}$ with such 
modulus squared any other one transformed by an  arbitrary 
real orthogonal matrix $\hat{L}_m$, $\hat{\omega'}=\hat{L}_m \hat{\omega}$, is equally valid mathematically.
 Then, we can always take the unitary transformation $\hat{\cu}_m$ 
such that the components of $\hat{\omega}$ are
\begin{align}
\label{140815.2}
\omega_i=|\widetilde{\gamma}_i|,~i=1,\ldots,m~.
\end{align}
 This is accomplished e.g.  by the diagonal unitary matrix
 $\hat{\cu}_m={\rm diag}(\frac{|\widetilde{\gamma}_1|}{\widetilde{\gamma}_1},\ldots,\frac{|\widetilde{\gamma}_m|}{\widetilde{\gamma}_m})$, with the 
understanding that if $\widetilde{\gamma}_i=0$ one takes 1 instead of $ \frac{|\widetilde{\gamma}_i|}{\widetilde{\gamma}_i}$ in  $\hat{\cu}_m$.
The main motivation to finally fix the action of  $\hat{L}_m$ 
 in this way (as a result $\hat{\ca}_m$ 
is invariant under the transformation of Eq.~\eqref{050815.1}) 
 is to avoid mixing between different channels in the resulting matrix of residues for the 
resonance, leaving intact 
the original strength (modulus) of the resonance coupling to each channel. Hence we can say that 
 the resonance signal around the pole is preserved  by the transformation (from the narrow resonance case 
it is well known that the phases of the residues are due to the
 background \cite{ref.130815.1,ref.300715.1,ref.060815.1}, cf. Sec.~\ref{sec.260316.5}).  
 In this way  the physical picture for the pole term in Eq.~\eqref{300715.1} as 
corresponding to the resonance exchange between channels driving by the original residues is preserved.

For the $T$ matrix corresponding to Eqs.~\eqref{140815.1} and \eqref{140815.2}  the residues are 
  $|\gamma_i|\eta_i$, with $\eta_i$ a pure phase factor given by $|\rho(s_P)_i|/\rho(s_P)_i$.
  However, the new  coefficients entering in the resulting 
SR, analogous to Eq.~\eqref{050815.2}, are not still positive definite because of 
 the complex numbers $dG(s_P)_i/ds$ and  $\eta_i$.
  We then allow by an extra unitary transformation  of the kind in Eq.~\eqref{050815.1}
 given by a  constant diagonal matrix $\hat{\cu}_m'={\rm diag}(e^{i\phi_1},\ldots,e^{i\phi_m})$, 
so that no mixing between resonance couplings to different channels arises.  
 The residues of the new $T$ matrix are  $|\gamma_i|e^{i\phi_i}\eta_i$, and the  phase factors $e^{i\phi_i}$  are fixed by requiring that
 the resulting partial compositeness coefficients, $X^R_i$, $i=1,\ldots,m$, are positive. Hence, 
\begin{align}
\label{050815.7}
X_i^R=&-e^{2i\phi_i}\eta_i^2 |\gamma_i|^2\frac{dG(s_P)_i}{ds}=|\gamma_i|^2\left|\frac{dG(s_P)_i}{ds}\right|~.
\end{align} 
  In terms of them one has the total compositeness, $X^R=\sum_{i=1}^n X_i^R$, and elementariness $Z^R=1-X^R$. 
 One should stress that these coefficients are entirely determined from the properties 
of the pole of the resonance ($s_P$ and $\gamma_i$) and no further ambiguity remains in the appropriate SR
 for a resonant state fulfilling {\bf I}, as the proper transformation of Eq.~\eqref{050815.1} has been fixed.
  This also implies that $X^R\leq 1$ should hold since the right SR  must be unique. At this point it is worth stressing that 
if the $T$ matrix were known in a region (no matter how small) of the physical $s$ axis  around $M_R^2$, 
 where the Laurent series Eq.~\eqref{300715.1} converges, one-variable complex analysis guarantees that  
the different contributions in this series  (in particular $s_P$ and $\gamma_i$)  
 could be determined unambiguous and model independently. 

 Let us stress that  we are relying on unitarity and analyticity, cf. Eq.~\eqref{140815.1},  and then
 our results could be applied to  a $T$ matrix with correct analytical properties 
in the complex $s$ plane, in particular along the unitarity cut. In this respect it is especially important 
to keep the whole energy dependence for $G(s)_i$ as dictated by Eq.~\eqref{040815.3}.\footnote{For its nonrelativistic reduction 
see Ref.~\cite{kang.090316.1}.}  In particular the restriction  $X^R\leq 1$ should be satisfied, 
 similarly as $s_P$, $\gamma$ and  $S_0$ are linked  to satisfy unitarity
  on the real axis wherever the Laurent series Eq.~\eqref{300715.1} converges. 
  It is clear that there is a close connection between 
narrow resonances (in which the two-body open channels are trapped for a relatively long time)
 and bound states, so that the compositeness relation for both cases should meet. Nevertheless, for a bound 
state both $\gamma_i^2$ and $-dG(s_P)_i/ds$  are positive and one does not need to take the absolute value 
in Eq.~\eqref{050815.7} \cite{ref.040815.2,ref.040815.3,ref.040815.4}.

\subsection{Narrow resonance case} 
\label{sec.260316.5}

We now assume that the resonance dominates 
the energy dependence of the  expansion in Eq.~\eqref{300715.1}
 on the real axis around $s=M_R^2$, 
 such that we can  take in good approximation  $S_0(s)$ as a constant matrix  
(this is typically the narrow resonance assumption). To simplify the writing we also drop the 
hat on top of the different matrices, so that  one should understand  in the following that the analysis is restricted to 
the set of $m$ open channels.

Implementing in the unitarity relation $S(s)S(s)^\dagger=1$ 
the expansion in Eq.~\eqref{300715.1} we have the equation
\begin{align}
&(s-s_P)(s-s_P^*)S_0 S_0^\dagger + (s-s_P)S_0 R^\dagger  + (s-s_P^*)R S_0^\dagger \nn \\
&+R R^\dagger=(s-s_P)(s-s_P^*)~.
\label{010715.7}
\end{align}
By identifying the coefficients of the different powers of $s$ we have the following set of equations
\begin{align}
\label{010715.8}
&S_0 S_0^\dagger=I~,\\
\label{010715.9}
&S_0 R^\dagger+RS_0^\dagger=0~,\\
\label{010715.10}
&s_P S_0 R^\dagger +s_P^*R S_0^\dagger - RR^\dagger=0~.
\end{align}
The first of them implies that $S_0$ is a unitary matrix.
One can find a family of solutions of these equations by writing 
\begin{align}
\label{010715.12}
S_0=&{\cal O}{\cal O}^T~,\\
\label{010715.11}
R=&i\lambda {\cal O} {\cal A} {\cal O}^T~,~\lambda \in \mathbb{R}~,
\end{align}
with ${\cal O}$ a unitary matrix, so that Eq.~\eqref{010715.8} is fulfilled.
 Although $S_0$ is symmetric the matrix ${\cal O}$  in Eq.~\eqref{010715.12} is not 
necessarily symmetric.\footnote{A particular solution would be 
${\cal O}=S_0^{1/2}$.} 
 We have from Eqs.~\eqref{010715.9} and \eqref{010715.10} that the matrix ${\cal A}$ satisfies, respectively,
\begin{align}
\label{090416.3}
i\lambda {\cal A}-i\lambda {\cal A}^\dagger=&0~,\\
\label{090416.4}
-2\Ima s_P {\cal A}+\lambda {\cal A}{\cal A}^\dagger=&0~.
\end{align}
From the first equation it is clear that 
\begin{align}
\label{010715.13}
{\cal A}^\dagger=&{\cal A}~,
\end{align}
while we can always take the normalization constant $\lambda$ such that additionally Eq.~\eqref{090416.4} is fulfilled 
 with  
\begin{align}
\label{010715.14}
{\cal A}^2=&{\cal A}~,\\
\label{010715.15}
\lambda=&2\Ima s_P=-2M_R\Gamma_R~.
\end{align}
As a result of Eqs.~\eqref{010715.13} and \eqref{010715.14} it follows that ${\cal A}$ is a projection operator. 
It is also a symmetric matrix, cf. Eq.~\eqref{010715.11}, 
and then its matrix elements must be real because it is a Hermitian matrix, cf. Eq.~\eqref{010715.13}.
 Additionally, ${\cal A}$ is a rank 1 projector operator by direct analogy between
 Eqs.~\eqref{300715.4} and \eqref{010715.11}.  We can  prove this conclusion explicitly here 
for the narrow resonance case by noticing that 
\begin{align}
\label{070416.1}
{\rm Tr}{\cal A}=&-\frac{1}{\lambda}{\rm Tr}|\widetilde{\gamma}||\widetilde{\gamma}|^T
=\frac{\sum_{i=1}^m |\gamma_i|^2 |\rho(s_R)_i|}{M_R\Gamma_R}=1~.
\end{align}
The equality to 1 of this sum stems from  the replacement of $\rho(s_R)\to \rho(M_R^2)$ in Eq.~\eqref{070416.1},
 since $\Gamma_R\ll M_R$, and the standard formula for the total decay width of a narrow 
resonance \cite{kang.090316.1}, 
\begin{align}
\label{090316.1}
\Gamma_R=&\sum_{i=1}^m \frac{|\gamma_i|^2 q(M_R^2)_i}{8\pi M_R^2}~.
\end{align}

Then, for the narrow resonance case we can write the $S$ matrix, Eq.~\eqref{300715.1}, around the resonance 
mass as
\begin{align}
\label{260316.5}
S(s)=&{\cal O}\left(I+\frac{i\lambda \ca}{s-s_P}\right){\cal O}^T~,
\end{align}
with $\ca$ a rank 1 projection operator and $\lambda$ given by Eq.~\eqref{010715.15}.
 For a given projection $\ca$ one can identify a resonance $S$ matrix, $S_R(s)$, given 
by the matrix between brackets in the previous equation, namely, 
\begin{align}
\label{260316.6}
S_R(s)=&I+\frac{i\lambda \ca}{s-s_P}~.
\end{align}
However, this identification is not unique because $\ca$ can be changed by the action of a real 
orthogonal matrix, as already discussed after Eq.~\eqref{260316.2}, $\ca\to L \ca$ and $\co\to \co L^T$. 
 Since $\co$ is  unitary  Eq.~\eqref{250316.5} necessarily holds and $L$ is finally fixed by the requirement 
of Eq.~\eqref{140815.2}, $\omega_i=|\widetilde{\gamma}_i|$. In this way, the resonance projection operator 
is determined and from this analysis it is clear that 
the phases of the original couplings $\gamma_i$ are due to the nonresonant terms through the unitary 
matrix ${\cal O}$, cf. Eqs.~\eqref{010715.12} and \eqref{260316.5}.\footnote{One has to take into account that 
$\rho(s_P)_i$, $i=1,\ldots,m$ are real in the narrow resonance limit $\Gamma_R/M_R\to 0$. \label{foot.090416.1}} 

Nevertheless, one has to make still one more transformation 
of the type of Eq.~\eqref{140815.1} to guarantee that the resulting compositeness coefficients $X_i$, $i=1,\ldots,m$, are real.  
 This extra transformation is accomplished by a  diagonal unitary matrix, 
analogously as done above in Eq.~\eqref{050815.7}, and is needed mainly because $dG(s_P)_i/ds$, $i=1,\ldots,m$ are
 complex, cf. footnote \ref{foot.090416.1}. 
In the transformed $S$ matrix, $S_u(s)$, the unitary matrix ${\cal O}$  transforms as ${\cal O}\to {\cal U}{\cal O}$,
 while the resonance $S$ matrix, $S_R(s)$, remains untouched.

In summary, our derivations in Secs.~\ref{sec.090316.1} and \ref{sec.260316.3} can be considered as
 the generalization of the narrow resonance case by 
extrapolating analytically in the resonance pole position $s_P$ from the real axis at $M_R^2$ (that corresponds to the 
limit of a narrow resonance $\Gamma_R/M_R\to 0$) up to 
$s_P$, along a contour that runs parallel to the imaginary $s$ axis.
 However, this extrapolation is only possible while {\bf I} 
holds. Among other facts notice that $dG(s)_m/ds$ has as branch-point singularity at $s_m$, and let us recall that 
$s_m<M_R^2<s_{m+1}$. 
If condition {\bf I} is not met one enters in a new qualitative physical picture that manifests by the 
fact that the Laurent series of Eq.~\eqref{300715.1} does not converge for any real $s$.

\begin{table*}[ht]
\begin{center}
\begin{tabular}{cc|cccc|cc} 
\hline
\hline
  Name of the states & Pole: $\sqrt{s_P}$ [MeV] &  $X^R_{\pi\pi}$ & $ X^R_{\bar{K}K}$ & $X^R_{\eta\eta}$  & $X^R_{\eta\eta'}$   & $X^R$ & $Z^R$  \\
$f_0(500)$ ~\cite{ref.070815.2} & $442^{+4}_{-4}-i 246^{+7}_{-5}$ & $0.40_{-0.02}^{+0.02}$ & $\cdots$ & $\cdots$ & $\cdots$ & $0.40_{-0.02}^{+0.02}$  & $0.60_{-0.02}^{+0.02}$  \\ 
$f_0(980)$ ~\cite{ref.070815.2} & $978^{+17}_{-11}-i 29^{+9}_{-11}$ & $0.02_{-0.01}^{+0.01}$ & $0.65_{-0.26}^{+0.27}$ & $\cdots$ & $\cdots$ & $0.67_{-0.27}^{+0.28}$ & $0.33_{-0.27}^{+0.28} $\\
 $f_0(1710)$ ~\cite{ref.100815.1}  & $1690^{+20}_{-20}-i 110^{+20}_{-20}$ & $0.00^{+0.00}_{-0.00}$ & $0.03^{+0.01}_{-0.01}$ & $0.02^{+0.01}_{-0.01}$ &  $0.20^{+0.07}_{-0.07}$   &
 $0.25^{+0.10}_{-0.10}$ & $0.75^{+0.10}_{-0.10}$ \\
$\rho(770)$ ~\cite{ref.070815.2} & $760^{+7}_{-5}-i 71^{+4}_{-5}$ &  $0.08_{-0.01}^{+0.01}$ & $\cdots$ &  $\cdots$ & $\cdots$ &  $0.08_{-0.01}^{+0.01}$  & $0.92_{-0.01}^{+0.01}$ \\    
\hline
                                  &                    & $X^R_{K\pi}$ &  &   &    & $X^R$ & $Z^R$   \\
$K_0^*(800)$~\cite{ref.070815.2}  & $643^{+75}_{-30}-i303^{+25}_{-75}$ & 0.94$_{-0.52}^{+0.39}$ &   &   &  & 0.94$_{-0.52}^{+0.39}$ & 0.06$_{-0.52}^{+0.39}$      \\ 
$K^*(892)$~\cite{ref.070815.2}  & $892^{+5}_{-7}-i25^{+2}_{-2}$ & 0.05$_{-0.01}^{+0.01}$ &  & &  & 0.05$_{-0.01}^{+0.01}$ & 0.95$_{-0.01}^{+0.01}$      \\
\hline
                        &        & $X^R_{\pi\eta}$ & $X^R_{\bar{K}K}$ & $X^R_{\pi\eta'}$  &     & $X^R$ & $Z^R$  \\
$a_0(1450)$~\cite{ref.070815.2} &  $1459^{+70}_{-95}-i174^{+110}_{-100}$ & 0.09$_{-0.07}^{+0.03}$ & 0.02$_{-0.02}^{+0.12}$ & 0.12$_{-0.09}^{+0.22}$  &  & 0.23$_{-0.18}^{+0.37}$ &  0.77$_{-0.18}^{+0.37}$   \\ 
\hline
    &    &  $X^R_{\rho\pi}$  &  &   &      & $X^R$ & $Z^R$   \\
  $a_1(1260)$~\cite{ref.180815.3}  & $1260-i250$  & 0.46 &  &    &    &  0.46 &  0.54 \\ 
\hline
Hyperon with $I=0$   &  & $X^R_{\pi\Sigma}$ & $X^R_{\bar{K}N}$ &  &    & $X^R$ & $Z^R$  \\
  $\Lambda(1405)$ broad~\cite{ref.130815.3}    & $1388^{+9}_{-9}-i114^{+24}_{-25}$ &  $0.73^{+0.15}_{-0.10}$ & $\cdots$ &  &   & $0.73^{+0.15}_{-0.10}$ & $0.27^{+0.15}_{-0.10}$ \\
 $\Lambda(1405)$ narrow~\cite{ref.130815.3}   & $1421_{-2}^{+3}-i\,19_{-5}^{+8}$   & 0.18$_{-0.08}^{+0.13}$ & 0.82$_{-0.17}^{+0.36}$ &  & &  1.00$_{-0.25}^{+0.49}$&  0.00$_{-0.25}^{+0.49}$ \\ 
\hline
Hyperon with $I=1$    &  & $X^R_{\pi\Lambda}$ & $X^R_{\pi\Sigma}$ & $X^R_{\bar{K}N}$ &    & $X^R$ & $Z^R$ \\
   Ref.~\cite{ref.130815.3}    & $1376_{-3}^{+3}-i\,33_{-5}^{+5}$  & 0.04$_{-0.00}^{+0.00}$ & 0.0$_{-0.0}^{+0.0}$ & $\cdots$  &   & 0.04$_{-0.00}^{+0.00}$& 0.96$_{-0.00}^{+0.00}$ \\
   Ref.~\cite{ref.130815.3}    & $1414_{-3}^{+2}-i\,12_{-2}^{+1}$  & 0.03$_{-0.00}^{+0.00}$ & 0.01$_{-0.00}^{+0.00}$ & 0.13$_{-0.04}^{+0.03}$ &   & 0.17$_{-0.04}^{+0.03}$ & 0.83$_{-0.04}^{+0.03}$ \\ 
\hline
 &  & $X^R_{D K}$ & $X^R_{D_s\eta}$ & $X^R_{D_s\eta'}$ &     & $X^R$ & $Z^R$   \\
 $D^*_{s0}(2317)$~\cite{Guo:2015dha} &$2321^{+6}_{-3}$  & 0.56$_{-0.03}^{+0.05}$ & 0.12$_{-0.01}^{+0.01}$  &   0.02$_{-0.01}^{+0.01}$
&     & 0.70$_{-0.05}^{+0.07}$ & 0.30$_{-0.05}^{+0.07}$     \\
\hline
 &   & $X^R_{J/\psi f_0(500)}$ & $X^R_{J/\psi f_0(980)}$  & $X^R_{Z_c(3900)\pi}$   & $X^R_{\omega\chi_{c0}}$   & $X^R$ & $Z^R$  \\
 $Y(4260)$~\cite{Dai,Ablikim:2013mio} & $4232.8 - i36.3$   & 0.00 & 0.02 & 0.02  &  0.17 &   0.21 & 0.79   \\ 
 \hline
 &   & $X^R_{\Sigma_c^+\pi^0}$ &   $X^R_{\Sigma_c^{++}\pi^-}$ &  $X^R_{\Sigma_c^0\pi^+}$   &   & $X^R$ & $Z^R$  \\
$\Lambda_c(2595)$~\cite{ref.200815.2} &  $2592.25 - i1.3$  & $0.11_{-0.02}^{+0.02}$  & $\cdots$   & $\cdots$   &  & $0.11_{-0.02}^{+0.02}$ & $0.89_{-0.02}^{+0.02}$  \\
\hline
\hline
\end{tabular}
\caption{Partial compositeness coefficients ($X_i^R$),  total compositeness ($X^R$) and elementariness ($Z^R$) for a set of resonances that satisfy condition {\bf I}. 
 \label{tab.100815.1}}
\end{center}
\end{table*}

\subsection{General $\co$ matrix}
\label{sec.090416.2}

When ${\bf I}$ does not hold we cannot conclude that $\hat{\cal O}$, 
cf.  Eqs.~\eqref{300715.4} and \eqref{140815.1}, corresponds to a unitary matrix 
 $\,\hat{\cu}_m^{-1}$. A further insight into this problem can be gained by writing a left polar 
decomposition for $\hat{\co}^{-1}$,
\begin{align}
\label{250316.1}
\hat\co^{-1}=Q{\cal V}~,
\end{align}
where ${\cal V}$ is a unitary matrix and $Q$ is positive definite Hermitian matrix.\footnote{Let us recall that $ \hat\co$ has inverse and this is why 
$Q$ is positive definite and not simply positive-semidefinite Hermitian matrix.} The matrix $Q$ can be further diagonalized as
\begin{align}
\label{250316.2}
Q=&{\cal Z}Q_d {\cal Z}^\dagger
\end{align}
where $Q_d={\rm diag}(d_1,\ldots, d_n)$ ($d_i>0$) and ${\cal Z}$ is a unitary matrix.
 In this form Eq.~\eqref{250316.1} and the  vector $\omega$ in Eq.~\eqref{250316.4} read
\begin{align}
\label{250316.3}
\hat \co^{-1}=&{\cal Z}Q_d{\cal Z}^\dagger{\cal V}~,\\
\omega=&{\cal Z}Q_d{\cal Z}^\dagger {\cal V} \widetilde{\gamma}~. 
\end{align}
For diagonal $\hat \co$, which is the one finally involved in Secs.~\ref{sec.260316.4} and \ref{sec.260316.5}, 
and it is also necessarily the case for the important  one-channel scattering, instead of Eq.~\eqref{250316.5} we have 
now
\begin{align}
\label{250316.6}
\hat{\omega}^T\hat{\omega}=\sum_{i=1}^m d_i |\widetilde{\gamma_i}|^2~.
\end{align}
 There is no way to fix the positive constants $d_i$ with just the information contained in the residue matrix ${\cal R}$ 
at the pole position $s_P$. However, when condition {\bf I} holds we can relate directly the SR at the pole position
 with a transformed unitary $S$ matrix  $\hat{S}_u(s)$, cf. Eq.~\eqref{140815.1},
 and conclude that $\hat{{\cal O}}_m$ is  unitary ($d_i=1$, $i=1,\ldots,m$).
 In other words, $d_i\neq 1$ cannot be excluded when calculating compositeness if 
${\bf I}$ is not met,\footnote{As we have explicitly checked in many examples.}
 and then $\hat{\cal O}$ does not correspond to the unitary-matrix-driven
 transformation of Eq.~\eqref{050815.1} from a resonance $S$ matrix.

\section{Application to some resonances of interest}  

We proceed to calculate the values of 
   compositeness and elementariness coefficients  from  Eq.~\eqref{050815.7} to examine the nature  
 of a set of resonances that  satisfy the condition  {\bf I} and whose pole position and residues   
are taken from literature. 
We should stress that our aim here is to exemplify our method for the calculation of $X_i$, cf. Eq.~\eqref{050815.7}, for several resonances of interest 
but  not to discern the goodness between possible different analyses/models for a given resonance nor  being
 exhaustive in listing such analyses.  
 We present  values for the partial compositeness coefficients  in  Table~\ref{tab.100815.1}, with the 
channels involved appearing as subscripts. 
  Several different types of resonances are considered, including  
light-flavor mesons, light-flavor baryons, and  mesons and baryons with heavy flavors. 

We briefly elaborate how the numbers in Table~\ref{tab.100815.1} are obtained. 
For a set of resonances   both the pole positions and the relevant 
residues (central values and uncertainties), are explicitly given in the references taken, which enable us to 
straightforwardly calculate the compositeness $X_i$ by using Eq.~\eqref{050815.7}. 
 The error bars given in Table~\ref{tab.100815.1} arise by propagating and adding in quadrature the errors from the pole position $s_P$ 
and the residues $\gamma_i^2$. 
 We do not attempt here to make an estimation of possible sources of systematic errors in the theoretical approach followed by that reference.
 Namely, the set of resonances to which we apply this procedure,  and related references used, are: 
 the light-flavor mesons $f_0(500)$, $f_0(980)$, $\rho(770)$, $K_0^*(800)$, $K^*(892)$, $a_0(1450)$ from Ref.~\cite{ref.070815.2}, 
the $f_0(1710)$ state~from Ref.~\cite{ref.100815.1}, 
the isoscalar and isovector hyperons from Ref.~\cite{ref.130815.3}, 
and the bound state $D^*_{s0}(2317)$ from Ref.~\cite{Guo:2015dha}.
For the $a_1(1260)$, we have taken into account the different normalization to  properly relate the coupling strength in Ref.~\cite{ref.180815.3}
 with $\gamma_i$ in this work.

 Now we consider the heavy-flavor states.  
For the light-flavor  resonances we observe that $Z^R$ is clearly dominant 
for the $\rho(770)$ and $K^*(892)$  (as it should be for genuine $q\bar{q}$ resonances \cite{pelaez,ref.070815.2}). 
 To a lesser extent this is also the case for the 
$f_0(1710)$ (identified in Ref.~\cite{ref.100815.1} as mainly a glueball state) and the 
$a_0(1450)$ (which is a representative of the second nonet
 of scalar resonances with  dominantly preexisting nature  \cite{ref.130815.2,ref.100815.1,ref.070815.2}). Regarding the 
lightest scalars we see that the two-body component is much more important,
 being overwhelmingly dominant for the $K^*_0(800)$ and 
 the largest for the $f_0(980)$. Nevertheless, the compositeness  for the $f_0(500)$ is more modest, 
although it raises  with increasing pion mass, so that $X^R\simeq 0.6$  for $M_\pi\simeq 220$~MeV  
(while still satisfying condition {\bf I}) \cite{ref.250815.2}. For the axial-vector $a_1(1260)$ we obtain that $X^R$
 and $Z^R$ are similar, which is compatible 
  with  Ref.~\cite{ref.250815.1} that finds an $N_c$ pole trajectory at odds with a $q\bar{q}$ resonance. 
  Regarding the light-flavor hyperon resonances,
 the $\Lambda(1405)$ is clearly a resonance mainly composed 
by the light-pseudoscalar mesons and the baryon octet, the lighter pole dominated  
by the $\pi\Sigma$ component and the heavier one by the $\bar{K}N$ component \cite{twopolenature}.
 However, these two-body components are small for the  
 two isovector-hyperon poles \cite{ref.040815.1,class}.

Notice that we have calculated the compositeness of the $f_0(980)$ in $K\bar{K}$ and that 
of the  $\Lambda(1405)$ in $\bar{K}N$, although {\bf I} is not really fulfilled because these channels have a threshold larger than $M_R^2$. 
 Our calculation in these cases is based on the fact that the difference 
between $M_R$ and  this threshold is significantly smaller than the width. 
 Thus, the transition through the threshold is  rather smooth  
 and then we can expect that $\omega_{m+1}\simeq |\widetilde{\gamma}_{m+1}|$ in  good approximation, 
cf. Eq.~\eqref{140815.2}, since the transformation above $s_{m+1}$ must be unitary.  As a result we can apply 
Eq.~\eqref{050815.7} as an approximation for the calculation of $X^R_{m+1}$ for this case.

 For the $Y(4260)$, we first estimate its quasi-two-body partial widths to $Z_c(3900)\pi$, 
$J/\psi f_0(500)$ and $J/\psi f_0(980)$ based on the analyses of  Refs.~\cite{Dai,Ablikim:2013mio}, while 
the decay width to $\chi_{c0}\omega$ is explicitly given in Ref.~\cite{Dai}. 
For the $Z_c(3900)\pi$ channel, it accounts  21\% of the  $J/\psi\pi\pi$ channel, according to Ref.~\cite{Ablikim:2013mio}. 
Regarding  the $h_c\pi\pi$ case, the possible important quasi-two-body channel is $h_cf_0(500)$, but then the interaction  
between $h_c$ and $f_0(500)$ is $P$-wave, in contrast to the $S$-wave interaction between $J/\psi$ and $f_0$ resonances. 
Therefore we assume that the rest of the decay widths for $Y(4260)$ are saturated by the quasi-two-body decays into 
$J/\psi f_0(500)$ and $J/\psi f_0(980)$, and then the data points with $m_{\pi\pi}<0.65$~GeV are 
identified as $J/\psi f_0(500)$ and the others as $J/\psi f_0(980)$~\cite{Ablikim:2013mio}. 
From the partial widths we estimate $|\gamma_i|$ and then $X^R_i$ can be calculated and given in Table~\ref{tab.100815.1}

Our results show that the two-body components are dominant  
for the $D^{*}_{s0}(2317)$ (a bound state in Ref.~\cite{Guo:2015dha}),  
although it also has  non-negligible elementariness. This conclusion is consistent 
with the large $N_c$ trajectory of its pole position \cite{Guo:2015dha}.
The residues for the $Y(4260)$ are estimated as discussed above. 
A small value of $X^R$ is found for $Y(4260)$, implying that this state is more like an elementary particle, 
a conclusion in agreement with Ref.~\cite{Dai} that uses the pole counting rule~\cite{Morgan}.

The $\Lambda_c(2595)$ was studied in Ref.~\cite{Hyodo13} in the isospin symmetric case.
However, its results are unstable under little changes of the isospin limit mass for pions, so that  we have made another
 delicate analysis \cite{Guo:2016wpy} and the results are stable and shown in  Table \ref{tab.100815.1}. 
 Because of isospin breaking in the threshold energies of the three $\pi\Sigma_c(2450)$ channels we can only calculate compositeness 
for the lightest one because then condition {\bf I} is satisfied. 
 Our robust conclusion is that the $\pi^0\Sigma_c^+$ component is small, 
with the rest of the channels (including $\pi^+\Sigma_c^0$ and $\pi^-\Sigma_c^{++}$) and components  
playing a dominant role in the resonance composition.

\section{Conclusions}

In this work we have derived a probabilistic interpretation of the compositeness relation
at the pole of a resonance with a convergent Laurent series for $s$ around ${\rm Re} s_P$
 (this is called condition {\bf I}), with $s_P$ the resonance pole position. 
It then allows one to calculate the compositeness coefficients for those channels that are open up to $s={\rm Re}s_P$.  
 The key point is to transform to a new unitary and symmetric $S$ matrix that shares the same strength in the resonance couplings but   
 has   a meaningful compositeness relation.  The narrow-resonance case is discussed  
in detail too. It is shown that the general process can be considered as the analytical extrapolation in $s_P$ from the narrow-resonance 
case (real $s$ axis) up to the final pole position $s_P$ along a contour parallel to the imaginary $s$ axis. 

We have shown that the compositeness relation is an expression of the proper normalization to 1 of a rank 1 projection operator 
associated to the pole state. It has also been demonstrated in full generality that elementariness for a lower number of channels 
considered includes the compositeness of other heavier channels. We have furthermore shown that there is indeed an infinity of possible 
compositeness relations, all them related by an orthogonal-like transformation of the $S$ matrix that is actually driven by a unitary matrix. 
Of these relations we can select the physically suited one if condition {\bf I} holds.  The narrow resonance case clearly shows that 
the role of the transformation that allows us to end with such expression is twofold. First, one can  
 isolate in this way a resonant $S$ matrix by removing 
nonresonant contributions.  Second, one can get rid of those phases associated with free particles in the asymptotic states, so as to 
end with a positive and real $X_i$.

The final expression for the compositeness $X_i$ is  $|\gamma_i^2\frac{dG(s_P)_i}{ds}|$, 
being  $\gamma_i^2$ the residue of the resonance to channel $i$
 and $G(s_P)_i$ the unitary two-point loop function for the same channel.  
 As stated, this expression does not hold for all types of resonance poles,
 unless they satisfy the condition {\bf I}. The latter can be stated as the resonance pole $s_P$ must lie
 in an unphysical Riemann sheet adjacent to the physical one in the region $s_m<s<s_{m+1}$, 
such that $s_m<{\rm Re} s_P<s_{m+1}$, being $s_m$ the threshold of channel $m$. Then, the previous 
expression can be applied to calculate $X_i$ for all the channels with $s_i\leq s_m$. If {\bf I} is not met 
one cannot exclude that $X_i$ is larger than one due to the appearance of extra real positive constants $d_i$, as it has been 
discussed.

 We have exemplified this method by calculating the two-body components for several resonances of interest, taking  
their pole position and residues from previous results in the literature. 
At this point we have to emphasize that we do not pretend to perform an exhaustive study of theoretical analyses/models 
in the literature for each resonance, but mainly using outputs from serious studies that already provide us with the necessary 
information to evaluate $X_i$. In this respect, the uncertainties in $X_i$ are estimated by  
propagating  errors inferred from the literature and added them quadratically, without any attempt to estimate systematic 
 uncertainties that could arise from the explicit models used to extract pole properties. 
 Further applications of this method to other resonances could provide important information.

\paragraph*{Acknowledgements} 
  We would like to thank E.~Oset and T.~Hyodo for interesting discussions.
 This work in supported in part 
by the MINECO (Spain) and ERDF (European Commission) grant FPA2013-40483-P and the Spanish
Excellence Network on Hadronic Physics FIS2014-57026-REDT,  
the National Natural Science Foundation of China (NSFC) under Grant Nos.~11575052 and 11105038, the Natural 
Science Foundation of Hebei Province with contract No.~A2015205205, 
the grants from the Education Department of Hebei Province under contract No.~YQ2014034, 
the grants from the Department of Human Resources and Social Security of Hebei Province with contract No.~C201400323.


\end{document}